\documentclass[acmsmall]{acmart}
\AtBeginDocument{%
  }

\renewcommand\footnotetextcopyrightpermission[1]{} %
\setcopyright{none}
\newcommand{\tool}{{PixLift}\xspace}

\newcommand{\eg}{{e.g.,}\xspace}
\newcommand{\ie}{{\it i.e.,}\xspace}

\setcopyright{none}
\renewcommand\footnotetextcopyrightpermission[1]{} %
\usepackage{subfigure}
\usepackage{xspace}

\begin{document}

\title{PixLift: Accelerating Web Browsing via AI Upscaling}

\author{Yonas Atinafu}
\affiliation{%
  \institution{New York University Abu Dhabi}
  \country{UAE}
}

\author{Sarthak Malla}
\affiliation{%
  \institution{New York University Abu Dhabi}
  \country{UAE}
}

\author{HyunSeok Daniel Jang}
\affiliation{%
  \institution{New York University Abu Dhabi}
  \country{UAE}
}

\author{Nouar Aldahoul}
\affiliation{%
  \institution{New York University Abu Dhabi}
  \country{UAE}
}

\author{Matteo Varvello}
\affiliation{%
  \institution{Nokia Bell Labs}
  \country{USA}
}

\author{Yasir Zaki}
\affiliation{%
  \institution{New York University Abu Dhabi}
  \country{UAE}
}

\renewcommand{\shortauthors}{Atinafu et al.}

\begin{abstract}
    Accessing the internet in regions with expensive data plans and limited connectivity poses significant challenges, restricting information access and economic growth. Images, as a major contributor to webpage sizes, exacerbate this issue, despite advances in compression formats like WebP and AVIF. The continued growth of complex and curated web content, coupled with suboptimal optimization practices in many regions, has prevented meaningful reductions in web page sizes. 
    This paper introduces \textit{\tool}, a novel solution to reduce webpage sizes by downscaling their images during transmission and leveraging AI models on user devices to upscale them. By trading computational resources for bandwidth, \tool enables more affordable and inclusive web access.  
    We address key challenges, including the feasibility of scaled image requests on popular websites, the implementation of \tool as a browser extension, and its impact on user experience. Through the analysis of 71.4k webpages, evaluations of three mainstream upscaling models, and a user study, we demonstrate \tool's ability to significantly reduce data usage without compromising image quality, fostering a more equitable internet.
\end{abstract}

\maketitle

\section{Introduction}
In many developing regions, internet access remains prohibitively expensive, with individuals constrained by limited data plans~\cite{LUMS_web_affordability, affordability_dev}. This digital divide hinders access to information, restricts economic opportunities, and exacerbates inequality~\cite{digital_divide, lythreatis2022digital}. Images are a major factor in the large sizes of modern webpages, accounting for a median of 40\% of total page weight on desktops and 37\% on mobile devices~\cite{http_archieve}. While advanced image formats like WebP~\cite{webp} and AVIF~\cite{avif}  have significantly improved compression efficiency, the overall size of webpages continues to grow due to increasingly complex and curated content~\cite{http_archieve}.

Previous research has explored innovative image optimization techniques, focusing on strategies such as fetching only a portion of the image (\eg 50\%) and reconstructing it via mirroring~\cite{browseLite}, reducing quality for progressive images~\cite{browseLite}, or even replacing images with semantically similar alternatives~\cite{webLego}.  Inspired by this body of work and the recent advances in generative AI~\cite{hassan2024, aldahoul2023, sesrm5, quicksrnet, srsubpixelcnn}, we propose a novel approach: request scaled down version of images on webpages during transmission and leverage an AI model running on a user’s device to scale them back up. This approach (\texttt{\tool}) trades local resources (GPU and CPU cycles, which are relatively cheap) for bandwidth (which is expensive and limited in developing regions). Despite its simplicity, this idea has several challenges addressed in this paper.

\vspace{0.05in}
\noindent
\textbf{Image Scaling Support.} The first challenge we address is assessing the availability of image scaling in the wild. Specifically, \textit{which percentage of web servers are capable of responding to modified requests for reduced image sizes?}. To answer this question, we conduct a large-scale crawl of the top 71.4k websites listed in the HTTP Archive~\cite{http_archieve}, testing various request strategies inspired by~\cite{browseLite}. We find that 10\% of these webpages have at least some partial support for remote image downscaling, with up to \textit{full} support for 1.5\% of the webpages. 

\vspace{0.05in}
\noindent
\textbf{Implementation Barrier.} While \tool could be implemented as a browser modification, such approach would hinder adoption by requiring users to switch browsers. Instead, we develop \tool as a Chromium extension, compatible with most desktop browsers and a few Chromium-based mobile browsers, such as Yandex and Kiwi. Additionally, numerous models exist for upscaling images. We integrate three mainstream super-resolution models (SESR-M5~\cite{sesrm5}, \textit{SR\_Sub-Pixel CNN}, and QuickSRNet Small 4X~\cite{quicksrnet}) with \tool and benchmark their performance. We find that ``QuickSRNet Small 4X'' is the most time efficient model, almost 10x faster than ``SR\_Sub-Pixel CNN''. However, this efficiency comes at an image quality reduction for about 40\% of the more challenging images. 

\vspace{0.05in}
\noindent
\textbf{User Experience.} While \tool aims to save bandwidth, its impact on resource usage and user experience remains a critical question. To evaluate this, we experiment with the 1,000 most popular Pakistani webpages under average mobile conditions for developing regions: 1.6 Mbps downlink/768 Kbps uplink with 150ms RTT~\cite{digital_divide}. We show that \tool, when assuming ``full'' remote image downscaling support, can save multiple MB of traffic for the majority of webpages. In turn, this makes webpages more responsive, with a median PLT reduction of 7 seconds. All this is achieved with minimal impact on CPU (10-20\% increase) and memory (1~GB), thanks to offloading image upscaling to the GPU. Even with low-end devices like the Galaxy A03s, upscaling up to 10 images -- which reside above the fold for less than 2\% of the tested webpages -- is completed before the browser \texttt{onload} event, thus not affecting user experience. This is accomplished while preserving high visual quality of images, as evident by a user study conducted with 100 participants.

\section{Related Work}
\label{sec:related}
Many researchers have investigated methods to enhance web browsing performance and user experience, but fewer have focused specifically on the role of images. We identify and briefly summarize three key studies in this area. WebLego~\cite{webLego} proposes an innovative solution that races semantically similar images, trading strict content fidelity for faster loading speeds. However, WebLego tends to increase the overall page size, making it unsuitable for developing regions with limited bandwidth. ScaleUp~\cite{newman2019scaling} is a browser extension that dynamically adjusts browser scaling to reduce the number of objects loaded above the fold. While ScaleUp is designed for low-bandwidth networks, its effectiveness is limited when websites load substantial content below the fold, as we observed in Pakistan (see Figure~\ref{fig:res:quality}). Moreover, ScaleUp can serve as a complementary approach to \tool. BrowseLite~\cite{browseLite} is a client-side tool that optimizes web images for data savings by opportunistically requesting alternative image encodings or partial image data. \tool builds on the ideas from BrowseLite by introducing a novel approach: requesting highly compressed images and leveraging local AI models for upscaling, further enhancing data savings and performance.

\section{Image Scaling Support} 
\label{sec:scaling} 

\vspace{0.05in}
\noindent
\textbf{Methodology.} We use BigQuery to analyze the HTTP Archive~\cite{harBigQuery}, querying data for the top 50,000 webpages (as per the site popularity in the Chrome User Experience Report (CrUX)~\cite{google_crux}). From these webpages, we extract the URLs of all embedded images, resulting in a comprehensive set of 3,185,097 unique images hosted by 54,754 unique domains. Inspired by BrowseLite's methodology~\cite{browseLite}, we analyze these unique image URLs to identify common patterns embedded in the URLs that might be related to image resizing, \eg ``width='', ``resize/(large or small or medium)'', ``300x300'', etc. This analysis identifies 35 regular expression patterns, 5 more than BrowseLite which though only analyzed 1,200 webpages.  To investigate whether these 54,754 domains ``actually'' support image resizing, we further select one random image URL per domain and test against these 35 regular expressions. 

\vspace{0.05in}
\noindent
\textbf{Results.} Our analysis reveals that among these 54,754 domains, 5\% support image resizing. The three most common resizing patterns are: ``\texttt{?width=d}'' (46.5\%), ``\texttt{?q=w\_d}'' (21.3\%), and ``\texttt{?\_dw}'' (17.4\%). To further evaluate image resizing capabilities, we analyzed the 50,000 websites from the HTTP Archive. Specifically, we quantify the proportion of images served by domains which support resizing across the initial 50,000 webpages. We find that 10\% of webpages contain images that can be resized. Among these, 15\% provide full resizing support for all images, while 50\% support resizing for only half of their images.

\section{\tool Design} 
\label{sec:design} 
To integrate super-resolution capabilities into a user's web browsing experience, we develop \tool as a Chromium extension (Manifest V3~\cite{manifestv3}). This approach simplifies deployment by avoiding browser modifications and facilitates easy installation on all Chromium-based desktop browsers, and few mobile browsers (\eg Yandex and Kiwi). The extension operates \textit{transparently}, processing images on-the-fly as users navigate webpages, and \textit{privately}, ensuring that no raw image data or sensitive metrics are left in the browser environment.

Given the goal of \tool to upscale images via super-resolution, we first integrate several of the most promising image upscaling models into the browser. To ensure compatibility and efficient deployment of the super-resolution models, we employ a three-stage conversion pipeline transforming these models from PyTorch (.pth)~\cite{pytorch} to TensorFlow.js (.tfjs)~\cite{tensorflowjs}:

\begin{enumerate}
    \item \textbf{Conversion to ONNX}: Export PyTorch models to the Open Neural Network Exchange (ONNX) format (.onnx)~\cite{onnx}, enabling interoperability between machine learning frameworks.
    
    \item \textbf{Conversion to TensorFlow}: Convert ONNX models to TensorFlow (.tf), integrating them within TensorFlow's ecosystem.
    
    \item \textbf{Conversion to TensorFlow.js (TFJS)}: Convert TensorFlow models to TensorFlow.js (.tfjs), including quantization to Float16 for improved performance and compatibility on mobile.
\end{enumerate}

\vspace{0.05in}
\noindent
\textbf{Model Conversion Process.} To ensure compatibility and efficient deployment of super-resolution models in the browser, we implement the following three-stage pipeline. First, we export the PyTorch (.pth) models~\cite{pytorch} to the Open Neural Network Exchange (ONNX) format (.onnx)~\cite{onnx} for interoperability. Second, we convert ONNX models to TensorFlow (.tf) format~\cite{tensorflowjs}. Finally, we transform TensorFlow models into TensorFlow.js (.tfjs)~\cite{tensorflowjs} with Float16 quantization~\cite{tensorflow_quantization_blog} for enhanced performance on mobile devices.

Currently, \tool supports three mainstream super-resolution models: a) SESR-M5~\cite{sesrm5}, a lightweight transformer-inspired model for efficient super-resolution on mobile platforms; b) SR\_Sub-Pixel CNN~\cite{srsubpixelcnn}, an optimized CNN using sub-pixel convolution for fast and accurate super-resolution; c) QuickSRNet Small 4X~\cite{quicksrnet}, which is suitable for real-time applications on less powerful devices.

\vspace{0.05in}
\noindent
\textbf{Environment Configuration.} Our solution is implemented as a browser-based extension leveraging TensorFlow.js for real-time enhancement. We configure TensorFlow.js to leverage WebGL~\cite{webgl} -- \texttt{await tf.setBackend(`webgl')} -- for GPU acceleration. We disable  \texttt{WEBGL\_FORCE\_F16\_TEXTURES} to balance  between performance and memory usage. Finally, we use \texttt{await tf.ready()} to confirm a successful model setup and deployment. 

\vspace{0.05in}
\noindent
\textbf{Model Loading and Validation.}

Models are loaded asynchronously using TensorFlow.js's\\ \texttt{tf.loadGraphModel(modelURL)}, a process designed to minimize blocking of the browser's main thread. During this loading phase, we record the initialization time as a metric to evaluate the model's loading efficiency and readiness. After the models are loaded, we perform a validation step to ensure compatibility with the extension's preprocessing and postprocessing pipelines. This includes verifying that the loaded model's input shapes align with the preprocessing output (e.g., tensors resized to \texttt{[128, 128, 3]}) and confirming that the output configurations meet the requirements for subsequent postprocessing stages. This validation ensures robust operation across diverse hardware and software environments, reducing the risk of runtime errors.

\vspace{0.05in}
\noindent
\textbf{Image Detection and Monitoring.} To detect images within a webpage, we use a \texttt{MutationObserver}~\cite{dom} that monitors the DOM for any additions or modifications to \texttt{<img>} elements, including changes to their \texttt{src} attributes. Detected images are queued for processing by the image upscaler model, ensuring comprehensive coverage of dynamically changing content. 
To optimize resource usage and to prevent performance degradation, we setup a concurrency limit ensuring that a model processes a maximum of two images concurrently (set by \texttt{MAX\_CONCURRENT = 2}). 
    
\vspace{0.05in}
\noindent
\textbf{Image Processing Pipeline.} Each detected image undergoes a 3-stage pipeline. The \textit{preprocessing} stage fetches the images as ``blobs'' and converts them to data URLs, a necessary step to ensure compatibility with the browser's rendering and TensorFlow.js input pipeline. This transformation allows images to be processed directly as tensors within the browser environment. Then, it transforms data URLs into tensors using the \texttt{tf.browser.fromPixels()} function. Next, it resizes the tensors using bilinear interpolation~\cite{gonzalez2018digital} to match the model input dimensions. Finally, it normalizes the pixel values to the range [0, 1], and aligns the inputs (normalized tensors) with model expectations through batching/transposing. Batching organizes multiple inputs for simultaneous processing, while transposing ensures the dimensions match the model's format.

During the \textit{inference} stage, preprocessed tensors are fed into the  \texttt{model.predict()} function to generate enhanced tensors. Finally, the \textit{postprocessing} stage transposes the output tensors back to their standard dimensions. Pixels are denormalized to [0, 255] values, tensors are converted back to data URLs and associated with their respective image \texttt{src} to ``upscale'' the current image with the enhanced one. 

\vspace{0.05in}
\noindent
\textbf{Quality Metrics Calculation: PSNR and SSIM.}
\begin{itemize}
    \item \textit{PSNR} (Peak Signal-to-Noise Ratio): Measures reconstruction fidelity between original and enhanced images using the Mean Squared Error (MSE) and a logarithmic formula.
    
    \item \textit{SSIM} (Structural Similarity Index Measure): Evaluates perceptual similarity focusing on luminance, contrast, and structural differences, computed on the CPU backend for consistency.
\end{itemize}

We use a custom bash script executed on our devices via Termux to monitor system resources usage during image processing. The script continuously records CPU and memory usage at one-second intervals, sourcing CPU data from \texttt{/proc/stat} and memory data from \texttt{/proc/meminfo}.

\vspace{0.05in}
\noindent
\textbf{Telemetry:} To facilitate \tool's performance evaluation, we include a telemetry module that collects several performance metrics to be reported to a PHP server (via HTTPS). When in use, it monitors the device CPU (\texttt{chrome.system.cpu.getInfo}), memory (\texttt{chrome.system.memory.getInfo}), and network usage\\(\texttt{chrome.webRequest}). Further, it measures the following metrics: 1) \textit{upscaling time}, or the total time for preprocessing, model inference, and postprocessing of an image; 2) \textbf{PSNR (Peak Signal-to-Noise Ratio)}: Measures the fidelity of the enhanced image compared to the original; and 3) \textbf{SSIM (Structural Similarity Index Measure)}: Evaluates perceptual similarity based on luminance, contrast, and structure.

\vspace{0.05in}
\noindent
\textbf{Implementation Workflow.} The extension operates as follows:
\begin{enumerate}
    \item \textbf{Configure Environment}: Set TensorFlow.js backend to WebGL and adjust precision settings.
    
    \item \textbf{Load Models}: Asynchronously load and validate selected super-resolution models using \texttt{tf.loadGraphModel(modelURL)}.
    
    \item \textbf{Observe DOM}: Use a MutationObserver to detect and queue new or modified \texttt{<img>} elements.
    
    \item \textbf{Process Images}: Preprocess tensors, perform model inference, and postprocess results for immediate display.
    
    \item \textbf{Enforce Concurrency}: Maintain a maximum of two concurrent processing tasks, dequeuing subsequent tasks as each completes.
    
    \item \textbf{Record Metrics}: Collect performance metrics (\textit{latency}, \textit{PSNR}, \textit{SSIM}, CPU, memory usage) and export them to the PHP server for analysis.
    
    \item \textbf{Cleanup and Error Handling}: Dispose of tensors, handle errors gracefully, and maintain state flags to ensure stable operations.
\end{enumerate}

\section{\tool Evaluation}
\label{sec:perf}
This section evaluates \tool with respect to: 1) user QoE quantified by image quality assessment and page load times, 2) resource usage measured via data, CPU/GPU, and memory utilization. 

\subsection{Methodology}
Our evaluation is composed of three parts: \textit{controlled} and \textit{in-the-wild} experiments, and a \textit{user study}. 

\vspace{0.05in}
\noindent
\textbf{Controlled.} We equip three testing devices with the Kiwi browser and the \tool addon. The three devices cover a range of entry-level to mid-range mobile phones~\cite{A03s,A12,A34}: \texttt{Galaxy A03s} an entry-level phone that costs \char`\~ \$100 with an Octa-core CPU (4x2.35 GHz Cortex-A53 \& 4x1.8 GHz Cortex-A53), a PowerVR GE8320 GPU, and 4GB RAM; \texttt{Galaxy A12} a low-end phone that costs \char`\~ \$130 with an Octa-core CPU (4x2.35 GHz Cortex-A53 \& 4x1.8 GHz Cortex-A53), a PowerVR GE8320 GPU, and 4GB RAM; and \texttt{Galaxy A34} a mid- to high-end phone that costs \char`\~ \$210 with an Octa-core (2x2.6 GHz Cortex-A78 \& 6x2.0 GHz Cortex-A55), a Mali-G68 MC4 GPU, and 4GB RAM. For each super-resolution model-device pairing, we conduct 500 experiments where we load 50 times a testing webpage we created that includes between 1 and 10 \textit{visible} images, \ie above the fold. The images are randomly chosen from a pool of 100 images with various sizes and image formats. 

\vspace{0.05in}
\noindent
\textbf{In-the-Wild.} For these experiments, we aim to test the performance of \tool with real websites \textit{today}, \ie given current support of image scaling (see Section~\ref{sec:scaling}) as we all as assuming \textit{full} support, \ie all served images can be requested with a smaller size. We develop a testing tool based on Puppeteer~\cite{puppeteer} to automate the loading of testing webpages via the Brave browser~\cite{brave}. We choose Brave because of its support for adblocking which removes noise from experiment runs. Our tool loads a testing webpage while intercepting all the requested images within a Page Load time (PLT)~\cite{MDN2024PageLoadTime}, which measures  the amount of time it takes for a webpage to fully load. For each image, it then performs local resizing while keeping the same compression quality. It then uses this data savings to derive potential size and time savings assuming \textit{full} \tool support. Next, it uses the strategy from Section~\ref{sec:scaling} to test whether such smaller images could already be requested \textit{today}, and adjust such savings. 

We build a dataset of the 1,000  most popular Pakistani webpages, identified from Tranco's top 1 million list~\cite{tranco} by filtering for webpages with the .pk domain. Each webpage is loaded five times with network conditions set to ``3G Fast'' (1.6 Mbps downlink/768 Kbps uplink with 150ms RTT)~\cite{3G_Fast}. This network condition reflects the ``average'' configuration based on Google's Lighthouse standard recommendation for mobile throttling, simulating the 85th percentile of mobile connection speeds~\cite{3G_Fast}. We emulate a typical modern viewport (393x852 pixels as in recent iPhone or Pixel phones), and assume a 200px downscaling, or roughly 50\% of the viewport; image height is dynamically adjusted to maintain image proportions. 

\vspace{0.05in}
\noindent
\textbf{User Study.} We randomly select 20 images from the 1,000 Pakistani websites, resize them to 200 pixels wide, and upscale them using SESR-M5, SR Sub-Pixel CNN, and QuickSRNet. We then ask 100 survey participants to compare the perceived quality of each of these 20 upscaled images with their original version. This study focuses on understanding user perceptions of super-resolution quality improvements, especially within a mobile browsing context.

\begin{figure*}[!htbp]
    \centering
    \subfigure[Upscaling time per model and mobile device.]{
        \includegraphics[width=0.45\textwidth]{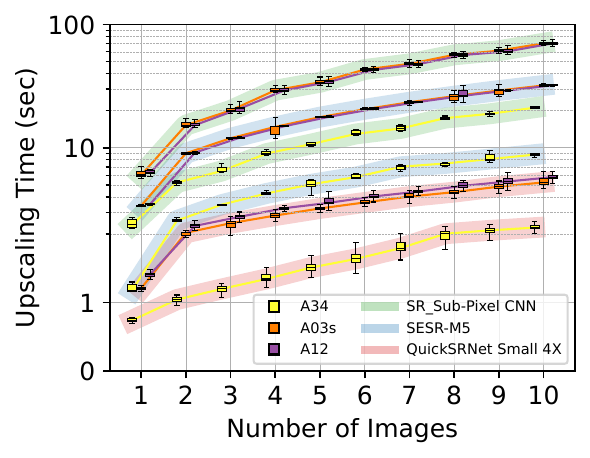}
        \label{fig:res:upscalingtime}
    }
    \subfigure[CPU memory (RAM) usage.]{
        \includegraphics[width=0.45\textwidth]{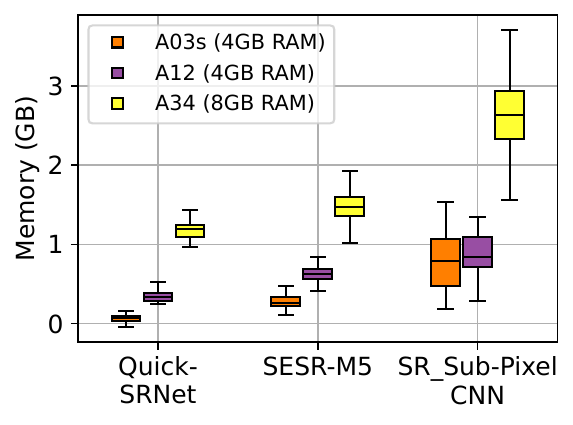}
        \label{fig:res:memory}
    }
    \subfigure[CDF of image quality scores for 20 images and 100 participants. Scores: 0 (much worse), 5 (same), 10 (much better).]{        \includegraphics[width=0.45\textwidth]{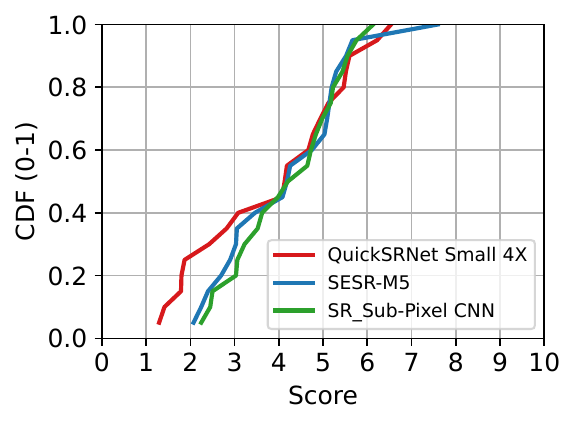}
        \label{fig:res:quality}
    }
    \caption{\tool performance evaluation: resource usage and visual quality assessment via user study.}
    \label{fig:res2}
\end{figure*}

\subsection{Results}

\vspace{0.05in}
\noindent
\textbf{Resource Usage and Image Quality.}
Figure~\ref{fig:res:upscalingtime} shows the total \textit{upscaling time} as a function of model, mobile device, and number of images located above the fold of a webpage. The upscaling time is mainly composed of the time required to transfer image data back and forth between the CPU and GPU, and the GPU processing time.  Each parameter combination in the figure is visualized as a boxplot referring to 50 runs. Given boxplots from different devices/models overlap, to improve the plot visibility we use a shade to indicate to which model a boxplot refers to. 

At a high level, the figure shows that ``QuickSRNet Small 4X'' is the most time-efficient model, almost 10x faster than ``SR\_Sub-Pixel CNN'' which is instead the least time-efficient model. Device-wise, the A03s and A12 perform quite similarly, while the A34 is, on average, 2x faster; \eg it can upscale 10 images in 2.5 seconds versus 5 and 6 seconds for the A03s and the A12, respectively. 

Figure~\ref{fig:res:upscalingtime} also shows a linear upscale duration growth between 1 and 2 images, followed by sublinear growth next. When considering a single image, the CPU-to-GPU transfer time is ``wasted,'' meaning that no GPU processing is possible at this time. The same is true with two images, given that we allow for two concurrent images to be handled by the GPU concurrently (see Section~\ref{sec:design}). However, if the GPU is powerful enough to handle two images concurrently, then the GPU processing time would drop. Indeed, the upscaling time lasts 1.2 seconds with 1 image and 2 seconds with 2 images, or a 16\% time reduction for both the A03s and the A12, whereas a 33\% time reduction is measured for the more powerful A34. As we increase the number of images to be upscaled past three, \tool's queuing mechanism ensures that a new image is transferred to the GPU as soon as an image is completed, while another image is being processed, thus amortizing the GPU transferring time. This behavior explains the sublinear growth realizing an average processing time of 0.2-0.6 seconds per image when considering \textit{batches} of 10 images, and variable GPU performance.

Figure~\ref{fig:res:memory} shows the CPU memory (RAM) usage as a function of device and model under test. Each boxplot refers to experiments with variable number of images, hence the variability. The plot shows that QuickSRNet has the lowest memory usage due to its efficient design with a simpler architecture, fewer parameters, and minimal intermediate data. Conversely, R Sub-Pixel CNN has the highest memory usage because its sub-pixel convolution layers significantly expand intermediate feature maps during pixel rearrangement. More powerful devices like the A34 further amplify memory usage by allocating more resources for computations, providing faster upscaling times as shown in Figure~\ref{fig:res:upscalingtime}. With respect to CPU usage, \tool is quite light only accounting for a 10-20\% CPU increase depending on model complexity. 

Finally, Figure~\ref{fig:res:quality} shows the Cumulative Distribution Function (CDF) of the average score per image for each of the 20 images and per super resolution model. The key takeaway is that for 50\% of the images with the highest scores (4 or higher), there are no significant score differences among models. However, for the remainder of the images with lower scores, ``QuickSRNet Small 4X'' achieves lower scores, \eg 40\% of scores lower than 3 versus  20\% for ``SR\_Sub-Pixel CNN.'' Given the 10x reduction in upscaling time achieved by  ``QuickSRNet Small 4X'' (see Figure~\ref{fig:res:upscalingtime}), we conclude that such quality reduction affecting a subset of the images might be tolerable by \tool users. Nevertheless, an interesting avenue of future work is to design a model selection algorithm capable of switching between super-resolution models based on the observed trade-off between upscaling time and image quality. 

\begin{figure*}[!htbp]
    \centering
    \subfigure[CDF of delta page load time (PLT), \ie, the difference between original PLT and PLT obtained via \tool.]{
        \includegraphics[width=0.45\textwidth]{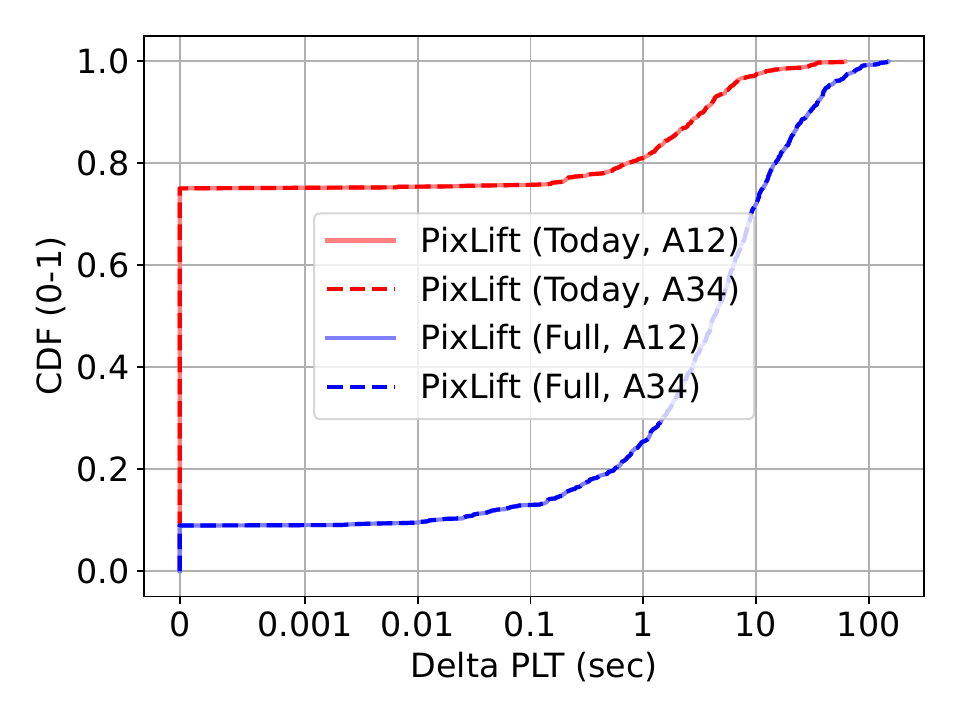}
        \label{fig:res:PLT}
    }
    \subfigure[CDF of delta page size, \ie, the difference between original webpage size and reduced webpage via \tool.]{
        \includegraphics[width=0.45\textwidth]{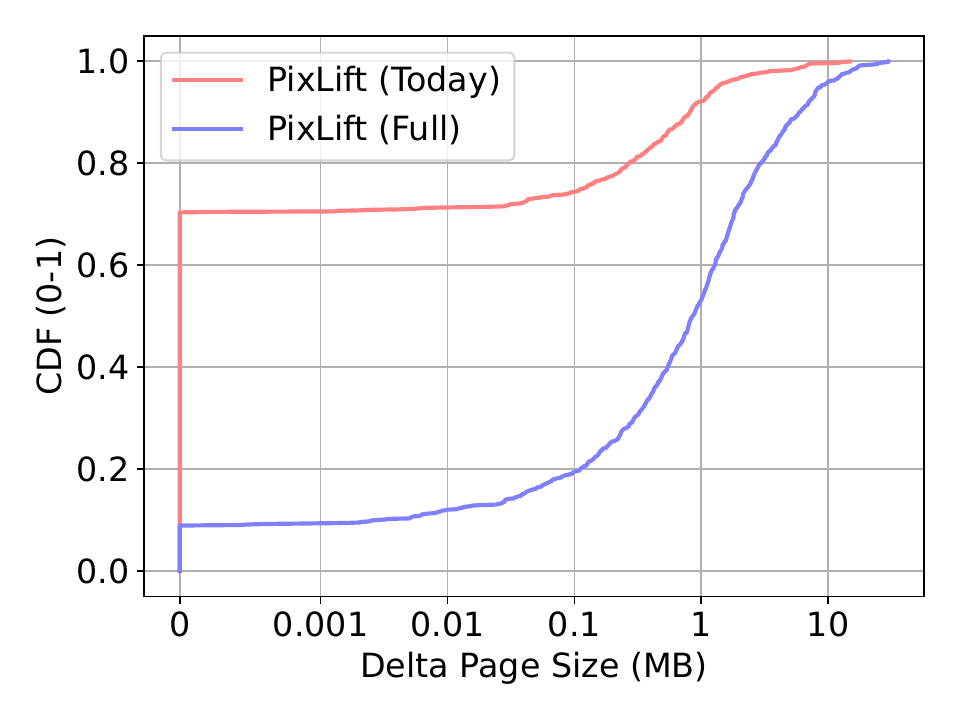}
        \label{fig:res:BDW}
    }
    \subfigure[CDF of image characteristics, \eg \textit{fetched} are all images transferred, and \textit{visible} are all images above the fold.]{
        \includegraphics[width=0.45\textwidth]{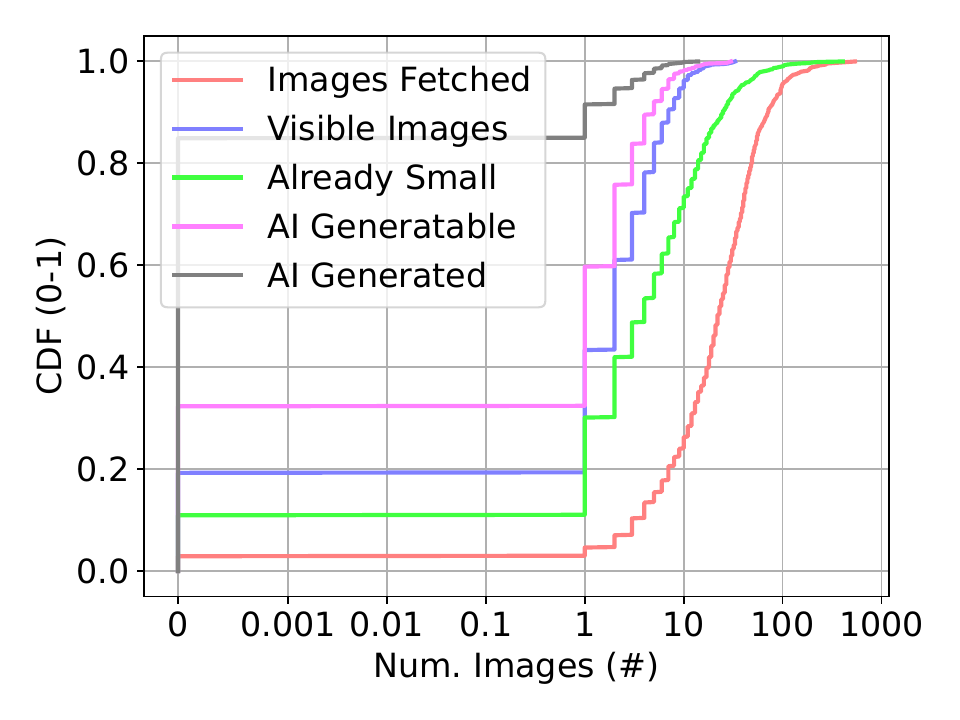}
        \label{fig:res:counters}
    }
    \caption{PixLift performance evaluation in the wild (1,000 most popular Pakistani webpages). \textit{Today} refers to current support for remote image downscaling; \textit{full} refers to assuming all images can be downscaled.}
\end{figure*}

\vspace{0.05in}
\noindent
\textbf{Performance.} Building on the previous results, and given the space limitations, we here investigate \tool performance assuming ``QuickSRNet Small 4X''. 
Figure~\ref{fig:res:PLT} shows several CDFs of \textit{delta} Page Load Time (PLT), \ie the difference between original PLT and PLT via \tool. We consider two versions of \tool: \textit{today}, which can only leverage downscaled images if currently supported by a server, and \textit{full}, which emulates full image downscaling support. For upscaling duration, we uses the empirical times from Figure~\ref{fig:res:upscalingtime} considering one low-end (A12) and one high-end device (A34).

Overall, Figure~\ref{fig:res:PLT} shows that \tool, when assuming full support, can significantly speedup the PLT by more than 7 seconds for the majority of webpages, and even tens of seconds for 30\% of the webpages. Slowdowns are rare (less than 7\% of webpages), suggesting that the GPU upscaling time rarely lasts longer than PLT. As expected from the analysis in Section~\ref{sec:scaling}, most pages (75\%) have no support for remote image downscaling. Neverthelees, 20\% of the tested Pakistani webpages can already be accelerated by multiple seconds today. Finally, these benefits are available for both low-end and high-end devices, suggesting that the extra time needed by low-end devices to upscale images can be absorbed within a PLT. These PLT savings are justified by the large data savings shown in Figure~\ref{fig:res:BDW}, \eg multiple MB for the majority of the webpages when considering full \tool support. 

Finally, Figure~\ref{fig:res:counters} shows several CDFs of image characteristics, \eg how many are \textit{fetched} versus \textit{visible}. The figure shows a large discrepancy between images fetched (22 at the median), and \textit{visible} (2 at the median) when assuming a modern viewport (393x852 pixels). While few of these images are \textit{already small} (4 at the median), \ie 200px wide or less, there is significant potential for \tool to downscale lots of images and thus reduce the overall data usage, as observed in Figure~\ref{fig:res:BDW}. Finally, the combination of \textit{visibile} images which are not \textit{already small} make up the set of images which are \textit{AI generatable} -- assuming ``full'' \tool support --  from which we can further derive the set of images which were actually \textit{AI generated} when intersecting with ``today'' image downscaling support. 

\section{Conclusion}

This work proposes \tool, a novel approach to reduce webpage sizes by downscaling images during transmission and leveraging AI models on user devices for upscaling, which can enable more affordable and inclusive web access. \tool trades computational resources for bandwidth, addressing the challenges of expensive data plans and limited connectivity in many developing regions.
In summary, the key contributions and findings of this study are:

\begin{itemize}
    \item Feasibility of Image Scaling: An analysis of 71.4k websites revealed that 10\% of webpages have partial support for remote image downscaling, with 1.5\% offering full support, indicating the potential to leverage existing server capabilities.
    \item Browser Extension Implementation: \tool is implemented as a Chromium extension, ensuring compatibility with most desktop browsers and some mobile browsers, operating transparently and privately.
    \item AI Upscaling Models: Three mainstream super-resolution models were evaluated, with ``QuickSRNet Small 4X'' demonstrating the best time efficiency, being almost 10x faster than ``SR\_Sub-Pixel CNN''. However, this efficiency comes at a quality reduction for about 40\% of the more challenging images.
    \item User Experience Improvement: \tool significantly reduces webpage loading times, with a median page load time (PLT) reduction of 7 seconds and even tens of seconds for 30\% of webpages when assuming full support for image downscaling. This is achieved with a minimal 10-20\% increase in CPU usage and 1 GB memory usage, offloading upscaling to the GPU.
    \item Performance on Diverse Devices: Experiments on a range of devices showed that \tool significantly speeds up PLT, and the extra time needed by low-end devices to upscale images can be absorbed within a PLT.
    \item Data Savings: \tool can achieve significant data savings by downscaling images, reducing the overall page size.
    \item Image Characteristics: The analysis of 1,000 Pakistani webpages showed a discrepancy between the number of images fetched (22 at the median) and the number of visible images (2 at the median), which suggests a large potential to reduce data usage by downscaling a large number of images.
    \item User Perception of Quality: A user study revealed that while 50\% of images showed no significant quality differences among models, 40\% of images showed lower quality scores when using the ``QuickSRNet Small 4X'' model.
\end{itemize}

In conclusion, \tool demonstrates the viability of trading computational resources for bandwidth to enhance web browsing, especially in areas with limited and expensive internet access. The ``QuickSRNet Small 4X'' model offers a practical balance between upscaling speed and image quality, making it suitable for real-world deployment. Further research could explore dynamic model selection based on the trade-offs between upscaling time and image quality. The results of this study suggest that \tool can significantly reduce data usage and improve page load times, contributing to a more equitable internet.

\bibliographystyle{ACM-Reference-Format}
\bibliography{sample-base}

\end{document}